\long\def\symbolfootnote[#1]#2{\begingroup%
    \def\thefootnote{\fnsymbol{footnote}}\footnote[#1]{#2}\endgroup} 

\def\aj{AJ}%
\def\apj{ApJ}%
\def\apss{Ap\&SS}%
\def\aap{A\&A}%
\def\mnras{MNRAS}%
\def\pasp{PASP}%
\def\pasj{PASJ}%
\newcommand{\ecritiso}{\mbox{$E_{\rm crit}^{\rm iso}$}}
\newcommand{\ecrittid}{\mbox{$E_{\rm crit}^{\rm tid}$}}

\newcommand{\thalf}{t_{1/2}}

\newcommand{\thalfiso}{\mbox{$t_{1/2}^{\rm iso}$}}
\newcommand{\thalftid}{\mbox{$t_{1/2}^{\rm tid}$}}

\newcommand{\tdis}{\mbox{$t_{\rm dis}$}}
\newcommand{\tdistid}{\mbox{$t^{\rm tid}_{\rm dis}$}}

\newcommand{\tcr}{t_{\rm cr}}
\newcommand{\trh}{t_{\rm rh}}
\newcommand{\trhi}{t_{\rm rh,i}}

\newcommand{\rh}{r_{\rm h}}
\newcommand{\rhi}{r_{\rm h,i}}
\newcommand{\rg}{R_{\rm G}}
\newcommand{\vg}{V_{\rm G}}

\newcommand{\rt}{\mbox{$r_{\rm t}$}}
\newcommand{\rj}{r_{\rm J}}
\newcommand{\rji}{r_{\rm J,i}}

\newcommand{\rv}{\mbox{$r_{\rm v}$}}

\newcommand{\tcc}{\mbox{$t_{\rm cc}$}}
\newcommand{\ncc}{\mbox{$N_{\rm cc}$}}
\newcommand{\rhcc}{\mbox{$r_{\rm h,cc}$}}

\newcommand{\nrh}{\mbox{$n_{\rm rh}$}}

\newcommand{\ve}{v_{\rm e}}

\newcommand{\dr}{\mbox{${\rm d}$}}

\newcommand{\rhoh}{\mbox{$\rho_{\rm h}$}}

\newcommand{\xie}{\mbox{$\xi_{\rm e}$}}
\newcommand{\xien}{\mbox{$\xi_{\rm e0}$}}

\newcommand{\dndt}{\mbox{$\dot{N}$}}
\newcommand{\dmdt}{\mbox{$\dr M/\dr t$}}
\newcommand{\dndteq}{\mbox{$\dot{N}$}}

\newcommand{\ko}{k}

\documentclass[useAMS,usenatbib]{mn2e}
\usepackage{amssymb,latexsym,graphicx,natbib,eufrak}
\newcommand{\rhj}{\mbox{$\mathfrak{R}$}}
\newcommand{\rhji}{\mathfrak{R}_i}

\newcommand{\rhjo}{\mathfrak{R}_1}
\newcommand{\rhjt}{\mathfrak{R}_2}

\newcommand{\ftj}{\mbox{$\mathfrak{F}$}}

\title[Lifetimes of star clusters]
  {Lifetimes of tidally limited star clusters with different radii}

\author[M. Gieles and H. Baumgardt]
  {M.~Gieles$^{1}$ and H.~Baumgardt$^2$\\
  $^1$ European Southern Observatory, Casilla 19001, Santiago 19, Chile \\
  $^2$ Argelander-Institut f\"{u}r Astronomie (Sternwarte), Universit\"{a}t  Bonn, Auf dem  H\"{u}gel 71, D-53121 Bonn, Germany 
}
\date{Released 2008 Xxxxx XX}

\pagerange{\pageref{firstpage}--\pageref{lastpage}} \pubyear{2006}

\def\LaTeX{L\kern-.36em\raise.3ex\hbox{a}\kern-.15em
    T\kern-.1667em\lower.7ex\hbox{E}\kern-.125emX}

\begin{document}         
\maketitle

   \begin{abstract} We study the escape rate of stars, $\dndt$, from
clusters with different radii on circular orbits in a tidal field
using analytical predictions and direct $N$-body simulations.  We find
that $\dndt$ depends on the ratio $\rhj\equiv\rh/\rj$, where $\rh$ is
the half-mass radius and $\rj$ the radius of the zero-velocity surface
around the cluster. For $\rhj\gtrsim0.05$, the ``tidal regime", there
is almost no dependence of $\dndt$ on $\rhj$. To first order this is
because the fraction of escapers per half-mass relaxation time,
$\trh$, scales approximately as $\rhj^{3/2}$, which cancels out the
$\rh^{3/2}$ term in $\trh$. For $\rhj\lesssim0.05$, the ``isolated
regime", $\dndt$ scales as $\rhj^{-3/2}$. The dissolution time-scale,
$\tdis$, falls in three regimes. Clusters that start with their
initial $\rhj$, $\rhji$, in the tidal regime dissolve completely in
this regime and their $\tdis$ is, therefore, insensitive to the
initial $\rh$. Our model predicts that $\rhji$ has to be
$10^{-20}-10^{-10}$ for clusters to dissolve completely in the
isolated regime. This means that realistic clusters that start with
$\rhji\lesssim0.05$ always expand to the tidal regime before final
dissolution. Their $\tdis$ has a shallower dependence on $\rhji$ than
what would be expected when $\tdis$ is a constant times $\trh$.  For
realistic values of $\rhji$, the lifetime varies by less than a factor
of 1.5 due to changes in $\rhji$. This implies that the ``survival" or
``vital" diagram for globular clusters should allow for more small
clusters to survive.  We note that with our result it is impossible to
explain the universal peaked mass function of globular cluster systems
by dynamical evolution from a power-law initial mass function, since
the peak will be at lower masses in the outer parts of galaxies. Our
results finally show that in the tidal regime $\tdis$ scales as
$N^{0.65}/\omega$, with $\omega$ the angular frequency of the cluster
in the host galaxy.
\end{abstract}

\section{Introduction}
\label{sec:intro}
Stars escape from clusters due to internal two and three body
encounters in which stars get accelerated to velocities higher than
the escape velocity. The resulting dissolution time-scale, $\tdis$,
depends on the number of stars, $N$, and the escape rate, $\dndt$, as
$\tdis\equiv -N/\dot{N}$. For a constant $\dndt$ the instantaneous
value of $\tdis$ is the remaining time to total dissolution.  For
clusters in isolation $\tdis$ scales linearly with the half-mass
relaxation time, $\trh$ \citep{amb38, 1940MNRAS.100..396S}.  The
presence of a tidal field speeds up $\dndt$ by roughly an order of
magnitude \citep{1961AnAp...24..369H, 1973ApJ...183..565S,
1997MNRAS.286..709G}.  When treating the tidal field as a radial
cut-off, as is often done in Fokker Planck and $N$-body simulations
\citep{1990ApJ...351..121C, 1997ApJ...474..223G, 2000ApJ...535..759T}
$\tdis$ also scales with $\trh$ (\citealt{2001MNRAS.325.1323B},
hereafter B01).
 
\citet{2000MNRAS.318..753F} demonstrated that for a realistic tidal field 
it is important to consider the finite time it takes stars to escape
through one of the Lagrange points. B01 showed that then
$\tdis\propto\trh^{3/4}$. This scaling was also found for more
realistic $N$-body simulations that include a stellar mass function
and stellar evolution (\citealt{1997MNRAS.289..898V,
2003MNRAS.340..227B}, hereafter BM03). BM03 only considered the
dependence of $\trh$ on cluster mass, $M$, since in their runs the
initial half-mass radius, $\rh$, and the initial tidal radius are
linked, so $\rh\propto M^{1/3}$.

Observations of young (extra-galactic) clusters show that the scaling
of $\rh$ with cluster mass, $M$, and galactocentric distance, $\rg$,
is considerably shallower: $\rh\propto M^{0.1}\,R^{\,0.1}_{\rm G}$
\citep{1999AJ....118..752Z, 2004A&A...416..537L, 2007A&A...469..925S}
than the Roche-lobe filling relation ($\rh\propto M^{1/3}\,R_{\rm
G}^{\,2/3}$), implying that massive clusters and clusters at large
$\rg$ are initially under-filling their Roche-lobe.

The relation between $\tdis$ and the median cluster radius, $\bar{r}$,
 was modelled by \citet{1971Ap&SS..13..300W} for clusters up to
 $N=250$. He found that $\tdis$ (in Myrs) scales with $\bar{r}^{3/2}$
 for $\bar{r}\lesssim0.01\,\rj$, with $\rj$ the Jabobi radius, being
 the radius of the zero-velocity surface around the cluster imposed by
 the tidal field. This is because for these clusters
 $\tdis\propto\trh$.  For $\bar{r}\gtrsim0.05\,\rj$ this scaling is
 not followed any more and $\tdis$ is shorter.  Large clusters are
 even more vulnerable to disruption when the effect of passing
 molecular clouds is included \citep{1958AJ.....63..465K,
 1985IAUS..113..449W, 2006MNRAS.371..793G}.

\citet{2005PASJ...57..155T} modelled the evolution of clusters of 
larger $N$ that are initially Roche-lobe under-filling using
collisionless $N$-body simulations. They confirm that for Roche-lobe
filling clusters $\tdis\propto\trh^x$, with $x=3/4$ as was found by
B01, and $x$ somewhat larger for Roche-lobe under-filling
clusters. They do not discuss the effect of radius on $\tdis$.  More
realistic $N$-body simulations, including various stellar initial mass
function and stellar evolution, of Roche-lobe under-filling clusters
were considered by \citet{1999PhDT.........1E}.  She concludes that
Roche-lobe under-filling clusters survive longer. However, this is for
a fixed $\rhi$ and different $\rj$. Since her simulations included
stellar evolution it is not possible to scale these results to the
same $\rj$.

Predictions for the survival probability of globular cluster, such as
the survival triangle \citep{1977MNRAS.181P..37F,
1997ApJ...474..223G}, rely on the assumption that $\tdis$ is a
constant times $\trh$. This is also an important assumption in a
recent attempt to explain the shape and the universality of the
turn-over of the globular cluster mass function
\citep{2007arXiv0704.0080M}. To be able to judge the applicability of
such models it is of importance that the relation between $\tdis$ and
$\rh$ is better understood for clusters with larger $N$.

The interplay between internal relaxation effects and external tidal
effects is the topic of this Letter. In \S~\ref{sec:runs} we present a
set of $N$-body simulations of clusters with different initial
Roche-lobe filling factors.  We introduce a simple analytical model
for $\dot{N}$ that includes internal dynamics and the external tides
in \S~\ref{sec:model}.  In \S~\ref{sec:sims} we confront our model
with the simulations and our conclusions are presented in
\S~\ref{sec:conclusions}.

\section{Description of the runs}
\label{sec:runs}

We simulate the evolution of clusters containing between $N=1024$á and
$N=32768$ particles, without primordial binaries and mass-loss by
stellar evolution, orbiting with angular frequency $\omega$ in a
steady point mass tidal field to simulate circular orbits in a
galactic potential.  The stellar masses are randomly drawn from a
power-law mass function with index $-2.35$ with the maximum mass 30
times larger than the minimum mass.

For the density distribution of the clusters we use
\citet{1966AJ.....71...64K} models, with $W_0=5$.  We define an
initial Roche-lobe filling factor as $\ftj\equiv\rt/\rji $, with
$\rji$ the initial Jacobi radius and \rt\ the King tidal radius, that
is, the radius where the stellar density of the
\citet{1966AJ.....71...64K} model drops to zero. We model four values
of $\ftj$, from 0.125 to 1.  The $W_0$ parameter and $\ftj$ set the
ratio of the initial $\rh$, $\rhi$, and $\rji$, which we denote by
$\rhji\,(\equiv\rhi/\rji)$.  Clusters with $N$=[1024, 2048, 4096,
8192] are run [16, 8, 4, 2] times to reduce statistical variations.
We also run an $N=4096$ simulation in isolation up to $\thalf$ to
determine the mass loss parameters.

All clusters are scaled to $N$-body units, such that $G = M = 1$ and
$E = -0.25$ \citep{1986LNP...267..233H}. Here, $E$ is the total
(potential and kinetic) initial energy of the cluster. In these units
the virial radius, $\rv\equiv GM/(-4\,E)$, equals unity and the
crossing time at $\rv$ is $2\sqrt{2}$.

The half-time, $\thalf$, is the when half of the initial number of
stars have become unbound, where bound is defined as the number of
stars within $\rj$.  We multiply $\thalf$ (in $N$-body times) by
$\omega$ to compare the results of different $\rhji$. The
dimensionless $\omega\thalf$ results are equivalent to physical times
for clusters at the same $\rg$.  A summary of the runs and the
resulting $\thalf$ and $\omega\thalf$ values is given in
Table~\ref{tab:runs}.

All $N$-body calculations were carried out with the \texttt{kira}
\citep{1996ApJ...467..348M, 2001MNRAS.321..199P} integrator on the
special purpose GRAPE-6 boards \citep{2003PASJ...55.1163M} of the
European Southern Observatory.

\begin{table}
\caption{Summary of the $N$-body simulations. }
\begin{center}
\begin{tabular}{rrrrrr}
\hline 

$N$  &$\ftj=\frac{\mbox{$\rt$}}{\mbox{$\rji$}}$&$\rhji=\frac{\mbox{$\rhi$}}{\mbox{$\rji$}}$& $\thalf$ & $\omega\thalf$& $\frac{\mbox{$\thalf$}}{\mbox{$\trhi$}}$       \\\hline

1024 & 1 &    0.186    & 114 &     7.20 &      5.8\\
1024 & 0.5 &    0.093    & 306 &     6.84 &     15.7\\
1024 & 0.25 &    0.047    & 768 &     6.07 &     39.4\\
1024 & 0.125 &    0.023    & 1718 &     4.80 &     88.2\\\vspace{0.03cm}
2048 & 1 &    0.186    & 174 &     11.0 &      5.1\\
2048 & 0.5 &    0.093    & 478 &     10.7 &     13.9\\
2048 & 0.25 &    0.047    & 1184 &     9.35 &     34.3\\
2048 & 0.125 &    0.023    & 2768 &     7.73 &     80.3\\\vspace{0.03cm}
4096 & 1 &    0.186    & 269 &     17.0 &      4.4\\
4096 & 0.5 &    0.093    & 765 &     17.1 &     12.4\\
4096 & 0.25 &    0.047    & 1833 &     14.5 &     29.6\\
4096 & 0.125 &    0.023    & 4235 &     11.8 &     68.5\\\vspace{0.03cm}
8192 & 1 &    0.186    & 435 &     27.5 &      3.9\\
8192 & 0.5 &    0.093    & 1188 &     26.5 &     10.6\\
8192 & 0.25 &    0.047    & 2925 &     23.1 &     26.1\\
8192 & 0.125 &    0.023    & 6614 &     18.5 &     59.0\\\vspace{0.03cm}
16384 & 1 &    0.186    & 670 &     42.3 &      3.3\\
16384 & 0.5 &    0.093    & 1867 &     41.7 &      9.1\\
16384 & 0.25 &    0.047    & 4705 &     37.2 &     22.9\\
16384 & 0.125 &    0.023    & 10456 &     29.2 &     51.0\\\vspace{0.03cm}
32768 & 1 &    0.186    & 1062 &     67.1 &      2.8\\
32768 & 0.5 &    0.093    & 3049 &     68.1 &      8.1\\
32768 & 0.25 &    0.047    & 8011 &     63.3 &     21.2\\
32768 & 0.125 &    0.023    & 17281 &     48.3 &     45.7\\
\hline
\end{tabular}
\label{tab:runs}
\end{center}
\end{table}
\begin{table}
\caption{Results of an isolated run. }
\begin{center}
\begin{tabular}{rrrrrr}
\hline 
       $N$  &$\nu$&$\tcc$& $\trhi$ & $\nrh$    &$\thalf$   \\\hline
4096 &0.081 &   209  & 68 & 3.1&  1.9$\times10^6$ \\
\hline
\end{tabular}
\label{tab:iso}
\end{center}
\end{table}

\section{Analytical model for the escape rate of tidally limited clusters}
\label{sec:model}

\subsection{The ``classical" Ansatz}

Lets first assume that a cluster, consisting of $N$ stars, loses a
constant fraction $\xie$ of its stars each $\trh$, so that we can
write for $\dndt$ (for example \citealt{1987degc.book.....S},
hereafter S87)

\begin{equation}
\dndteq=-\xi_e \frac{N}{\trh},
\label{eq:dndt}
\end{equation}
where $\trh$ is conventionally  expressed as \citep{1971ApJ...164..399S}

\begin{equation}
\trh=0.138\frac{N^{1/2}\rh^{3/2}}{\sqrt{\bar{m}G}\ln\Lambda},
\label{eq:trh}
\end{equation}
where $\Lambda=\gamma N$ and $\gamma=0.11$
\citep{1994MNRAS.268..257G}, $G$ is the gravitational constant and
$\bar{m}$ is the mean stellar mass.  The crossing time at $\rh$,
$\tcr$, is given by

\begin{equation}
\tcr=\ko\left(\frac{\rh^{3}}{GM}\right)^{1/2},
\label{eq:tcr}
\end{equation}
where $\ko$ is a constant of order unity depending on the cluster density profile.

The escape energy of stars in an isolated cluster, \ecritiso, is four
times the mean kinetic energy of stars in the clusters, so
$\ecritiso=0.8GM/\rh$ (S87) and the escape velocity, $\ve$, scales
with the stellar root mean square velocity  in the cluster, $v_{\rm rms}$, as
$\ve^2=4\,v^2_{\rm rms}$. From integration over a Maxwellian velocity
distribution the fraction of stars with $v^2>4\,v^2_{\rm rms}$ can be
determined. We refer to this escape fraction for isolated clusters as
$\xien$ and S87 showed that $\xien=0.0074$.

\subsection{The escape fraction as a function of cluster radius}
\label{ssec:xie}

When a cluster evolves in a tidal field the critical energy for escape is
\begin{equation}
\ecrittid=-\frac{3GM}{2\rj}.
\label{eq:ecrittid}
\end{equation}
For a point-mass galaxy, $\rj$ depends on $\omega$ and $M$ as
\begin{equation}
\rj=\left(\frac{G}{3\omega^2}\right)^{1/3}\,M^{1/3},
\label{eq:rj}
\end{equation}
where $\omega\equiv\vg/\rg$, with $\vg$ the circular velocity.  A
large $\omega$ means a strong tidal field, which results in a small
$\rj$.

The ratio $\ecrittid$ and $\ecritiso$ gives the relative reduction of
the escape energy due to the tidal field (following S87)

\begin{equation}
\Gamma=\frac{\ecrittid}{\ecritiso}=-\frac{3GM}{2\rj}{\mbox{\LARGE{/}}}\frac{0.8GM}{\rh}=-\frac{15}{8}\rhj,
\label{eq:gamma}
\end{equation}
where we have used $\rhj\equiv\rh/\rj$. Note that this $\Gamma$ is a
factor 1.5 higher than the original definition in S87, since his
result was based on $\ecrittid=GM/\rj$, which he later refines to
equation~(\ref{eq:ecrittid}).  We calculate $\xie$ as a function of
$\rhj$ by numerically integrating Maxwellian velocity distributions
for different $\rhj$ to determine the fraction of stars with
velocities $v^2\ge4\,[1-\Gamma(\rhj)]
v^2_{\rm rms}$.

\begin{figure}
\begin{center}
   \includegraphics[width=8.cm]{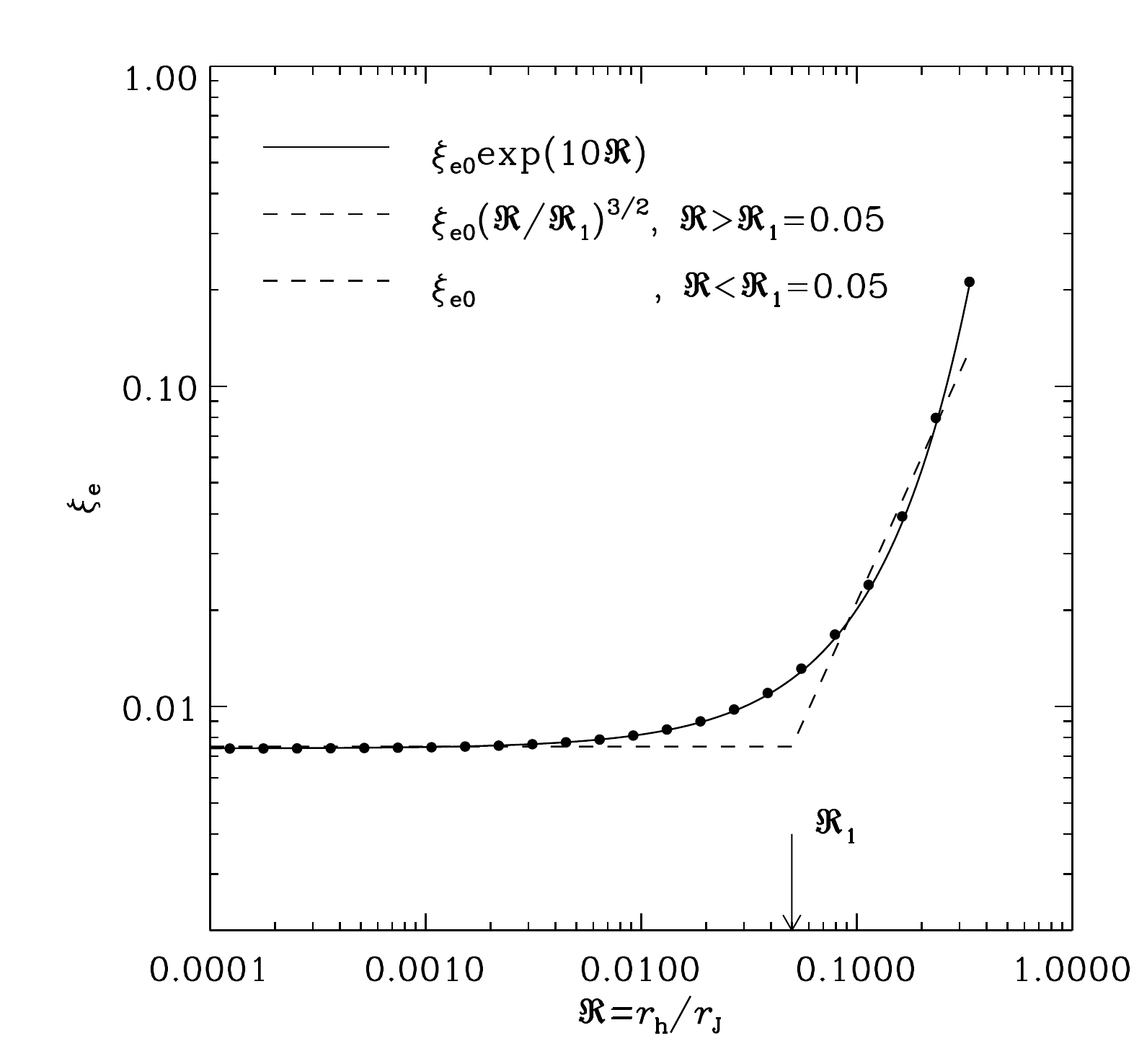} 
   \end{center} 
   \caption{The fraction of escapers, $\xie$, for different $\rhj$ for
   clusters in a tidal field. The points show the result of a
   numerical integration, the full line an exponential approximation
   and the dashed line a double power-law approximation.  }
   \label{fig:xi}
\end{figure}

In Fig.~\ref{fig:xi} we show that $\xie$ increases exponentially for
increasing $\rhj$ and can be well approximated by $\xien\exp(10\rhj)$.
For $\rhj\gtrsim 0.05$, the expression for \xie\ scales approximately
as $\rhj^{3/2}$, which has important consequences for $\dndt$ (see
equations~(\ref{eq:dndt})\,\&\,(\ref{eq:trh})).  We approximate $\xie$
by

\begin{eqnarray}
\xie&=&\xien, \hspace{2.5cm}\rhj<\rhjo \nonumber\\
       &=&\xien\left(\frac{\rhj}{\rhjo}\right)^{3/2},\hspace{1.0cm}\rhj\geq\rhjo,
\label{eq:xie}
\end{eqnarray}
where  $\rhjo=0.05$ is the boundary between the ``isolated regime" ($
\rhj\leq\rhjo$) and the ``tidal regime" ($\rhj>\rhjo$).

Substituting equations~(\ref{eq:trh}) \& (\ref{eq:xie}) in
equation~(\ref{eq:dndt}) and using equation~(\ref{eq:rj}) we find for
$\dndt$ in the tidal regime

\begin{equation}
\dndteq=-\left(\frac{\sqrt{3}\xien\ln\Lambda}{0.138\,\rhjo^{3/2}}\right)\,\omega.
\label{eq:dndt2}
\end{equation}
So $\dndt$ is independent of $\rh$ in the regime where
$\xie\propto\rhj^{3/2}$. This is because for a smaller(larger) star
cluster the shorter(longer) $\trh$ is balanced by the lower(higher)
$\xie$\symbolfootnote[2]{Ivan King noticed a remarkable similarity
between this result and equation~(53) in his 1966 paper. There he
shows that the escape rate of stars from a Roche-lobe filling cluster
is independent of position within the cluster. This is probably
because of the same physical reason, but he derived it in a different
way. }.  Equation~(\ref{eq:dndt2}) also shows that $\dndt$ depends
only marginally on $N$ through $\ln\Lambda$.

\subsection{Including the escape time}
\label{ssec:tesc}

\citet{2000MNRAS.318..753F} consider the time-scale of escape for stars 
in a cluster evolving in a tidal field. This time-scale is non-zero
because stars with energies (slightly) larger than the escape energy
still need a finite time to find one of the Lagrange points, where the
escape energy is lowest, to leave the cluster.

B01 derives an expression for $\dndt$ of stars in the potential
escaper regime, that is, with energies higher than the escape energy,
but still trapped in the potential, and shows that it scales as
$N\trh^{-3/4}\,\tcr^{-1/4}$, instead of $N\,\trh^{-1}$
(equation~\ref{eq:dndt}).  We include $\xie$ and define
\dndt\,analogous to equation~(\ref{eq:dndt}), as

\begin{equation}
\dndteq= -\xie\, \frac{N}{ \trh^{3/4}\,\tcr^{1/4}}.
\label{eq:dndtesc}
\end{equation}
With the expressions for $\xie$, $\trh$ and $\tcr$ we then find for
$\dndt$ in the tidal regime, including the escape time,
\begin{equation}
\dndteq=-\left(\frac{\sqrt{3}\xien}  {\ko^{1/4}\,\rhjo^{3/2}}\right)\left(\frac{\ln\Lambda}{0.138}\right)^{3/4}\,N^{1/4}\,\omega.
\label{eq:dndt3}
\end{equation}
From a comparison between equation~(\ref{eq:dndt2}) and
equation~(\ref{eq:dndt3}) we see that $\dndt$ becomes $N$ dependent
when we include the escape time, but is still independent of $\rh$.
The dissolution time-scale in the tidal regime, which we define as
$\tdistid\equiv -N/\dndt$, is then

\begin{equation}
\tdistid= A\,\left(\frac{N}{\ln\Lambda}\right)^{3/4}\frac{1}{\omega},
\label{eq:tdistid}
\end{equation}
with $A=0.138^{3/4}\,\rhjo^{3/2}\ko^{1/4}/(\sqrt{3}\xien)$. For a
$W_0=5$ cluster $\ko=3.85$, which together with the values for $\rhjo$
and $\xien$ from \S~\ref{ssec:xie} results in $A=0.277$.  The term
$(N/\ln\Lambda)^{3/4}$ can be well approximated by $B\,N^{\eta}$, with
$\eta\simeq0.6$ \citep{2005A&A...429..173L}. The values of $B$ and
$\eta$ depend slightly on the value of $\gamma$ in $\Lambda=\gamma
N$. We find the best agreement with the Roche-lobe filling simulations
for $\gamma=0.2$, which results in $B=0.5$ and $\eta=0.65$, so we can
write $\tdistid=0.138\,N^{0.65}/\omega$.  This scaling of $\tdis$ with
$N$ was also derived from observations \citep{2003MNRAS.338..717B,
2005A&A...429..173L,2005A&A...441..949G}. To compare the model to the
results of the simulations we derive $\thalf$ in the tidal regime,
$\thalftid$, from $\tdistid$. \citet{2005A&A...441..117L} show that
when $\tdis=B\,N^\eta$, with $B$ a constant, then
$\thalf=(B\,N^\eta/\eta)\,(1-[1/2]^\eta)$, so

\begin{equation}
\thalftid=0.077\,\frac{N^{0.65}}{\omega}.
\label{eq:thalftid}
\end{equation}
We assume that the effect of the escape time is the same for all
$\rhj>\rhjo$, so that we can apply equation~(\ref{eq:thalftid}) in
this regime. With these relations we also assume that $\dndt$ in the
pre-collapse and post-collapse phase is the same.

\subsection{Dissolution in the isolated regime}

We assume that clusters with $\rhj<\rhjo$ evolve in the same way as
clusters in isolation, up to the moment that $\rhj$ becomes equal to
$\rhjo$. This assumption is justified by equation~(\ref{eq:gamma})
from which we see that for a cluster with $\rhj=\rhjo$ the relative
contribution of the tidal field to the escape energy is less then
10\%.  In the absence of primordial binaries and mass-loss by stellar
evolution $\rh$ and $N$ remain roughly constant up to the moment of
core collapse, $\tcc$ (\citealt{2002MNRAS.336.1069B}, hereafter B02).
After $\tcc$, isolated clusters evolve in a self-similar way ($\trh
\propto t$), which results in two fundamental relations for the
evolution of $N$ and $\rh$ (\citealt{1984ApJ...280..298G}; S87; B02):

\begin{equation}
N(t)=\ncc\left(\frac{t}{\tcc}\right)^{-\nu}
\label{eq:nt}
\end{equation}
\begin{equation}
\rh(t)=\rhcc\left(\frac{t}{\tcc}\right)^{\frac{2+\nu}{3}},
\label{eq:rht}
\end{equation}
where $\rhcc$ and $\ncc$ are $\rh$ and $N$ at $\tcc$.  More
complicated relations, assuming a non-zero origin for these relations,
exist \citep{1994MNRAS.270..298G}.  Core collapse happens after a
multiple number of the initial $\trh$, $\trhi$: $\tcc=\nrh
\trhi$.

From equation~(\ref{eq:nt}) we find that $\thalf$ for clusters
evolving completely in the isolated regime, $\thalfiso$, is
\begin{equation}
\thalfiso=\left(\frac{0.138\times2^{1/\nu}\,\nrh}{\sqrt{3}}\right)\frac{N}{\ln\Lambda}\frac{\rhji^{3/2}}{\omega}.
\label{eq:thalfiso}
\end{equation}
Because clusters expand after $\tcc$, not all clusters that start with
$\rhji<\rhjo$ will reach $\thalfiso$ before they reach $\rhjo$.The
maximum $\rhji$ for which equation~(\ref{eq:thalfiso}) applies is
found from equations~(\ref{eq:rht}) \& (\ref{eq:thalfiso}) and depends
on $\rhjo$ and $\nu$ as

\begin{equation}
\rhjt=\rhjo\,\left(\frac{1}{2}\right)^{\frac{2+2\nu}{3\nu}},
\label{eq:rhjt}
\end{equation}
which for $\rhjo=0.05$ (Fig.~\ref{fig:xi}) and $\nu=0.05-0.1$ (B02)
results in $\rhjt\simeq6\times10^{-5}-6\times10^{-3}$. If we define
complete dissolution as the moment where only 5\% of the original
number of stars is still bound, then the corresponding value of
$\rhjt$ reduces to $\sim10^{-20}-10^{-10}$. This implies that
realistic clusters never dissolve completely in the isolated regime.

\subsection{Combining the isolated and the tidal regime}
\label{ssec:all}

Clusters that start with $\rhjt<\rhji<\rhjo$ evolve partially in the
isolated regime and partially in the tidal regime. Though our model
allows to numerically compute $\thalf$ for clusters in this regime, we
simply connect $\log\left[\omega\thalfiso(\rhjt)\right]$ and
$\log\left[\omega\thalftid(\rhjo)\right]$ with a straight line. The
slope of this line, representing the $\rhji$ dependence of
$\omega\thalf$, is
$\log\left[\thalftid(\rhjo)/\thalfiso(\rhjt)\right]/\log\left[\rhjo/\rhjt\right]$. This
slope is slightly $N$-dependent, and by using
equations~(\ref{eq:thalftid})\,\&\,(\ref{eq:thalfiso}) and the
parameters from the isolated run (Table~\ref{tab:iso}) we find that it
decreases from $0.55$ to $0.35$ between $N=1024$ and $N=32768$. For
$N=10^6$ the slope would be $0.15$.

\begin{figure}
\begin{center}
   \includegraphics[width=8.cm]{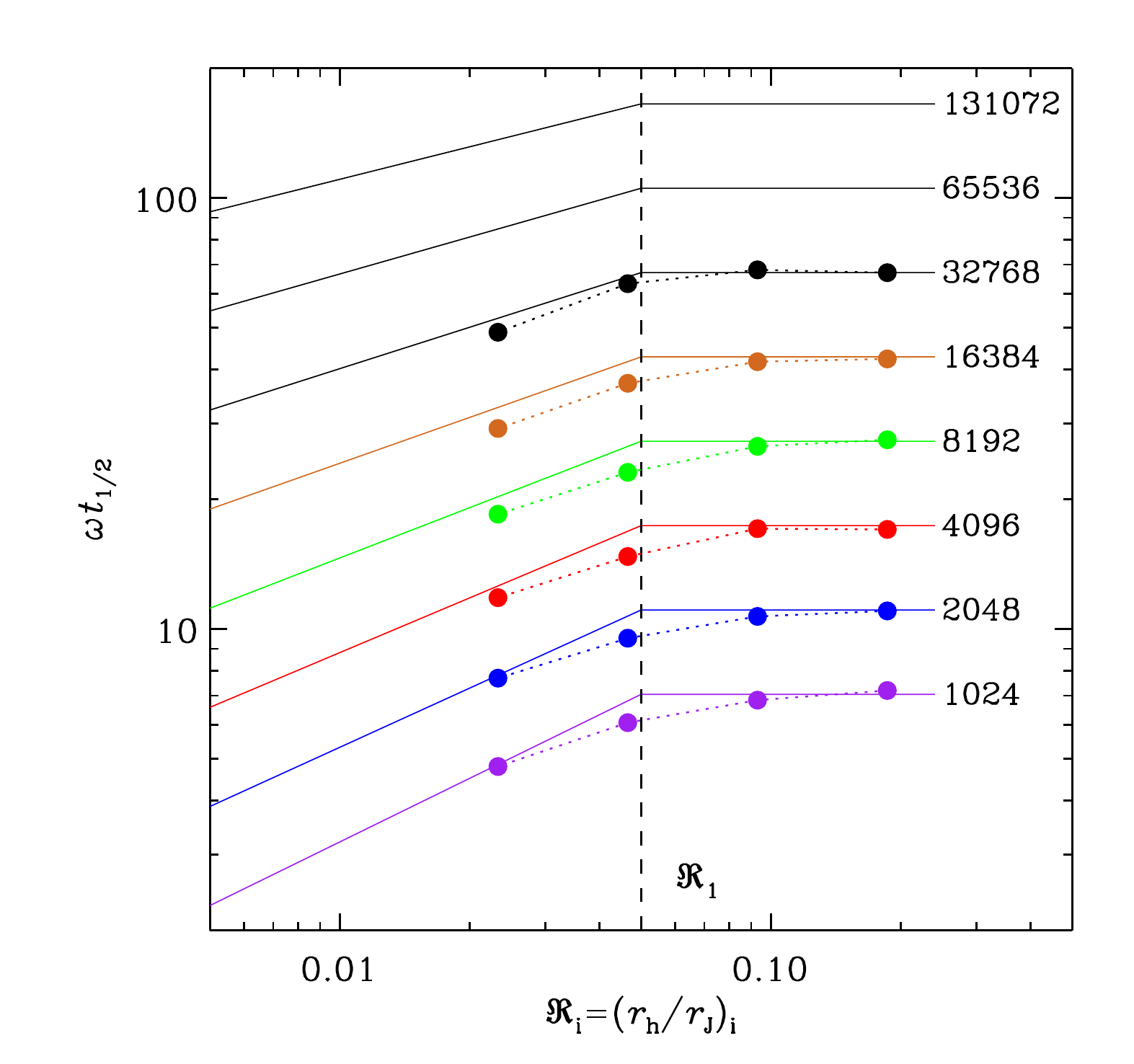}
   \end{center}
   \caption{Prediction for the dimensionless half-time,
      $\omega\thalf$, for clusters in the tidal regime
      (equation~\ref{eq:thalftid}) and part of the intermediate regime
      (\S~\ref{ssec:all}) for different $N$. Results of the
      simulations are shown as dots.  } \label{fig:model}
\end{figure}

In Fig.~\ref{fig:model} we show the $\omega\thalf$ following from our
model for different $N$ and a range of two orders of magnitude around
$\rhjo$. The results of the simulations are shown as dots and are
discussed in the next section.

\section{Comparison to $N$-body simulations}
\label{sec:sims}
The results for $\omega\thalf$ of the simulations of clusters with
different $N$ and $\rhji$ are presented in Fig.~\ref{fig:model}.  The
$\omega\thalf$ results for $\ftj=0.5$ clusters are nearly the same as
those for $\ftj=1$. For $\ftj=0.25(0.125)$ the $\omega\thalf$ values
are approximately 10\%(25\%) shorter compared to the Roche-lobe
filling results, whereas a scaling with $\trh$ predicts a difference
of a factor of $4^{1.5}(8^{1.5})\simeq8(23)$. From
Table~\ref{tab:runs} we see that clusters that start Roche-lobe
under-filling, have evolved for a much larger number of relaxation
times than Roche-lobe filling clusters by the time they reach
$\thalf$.

\section{Conclusions}
\label{sec:conclusions}
The dissolution time-scale of clusters evolving in a tidal field, in
dimensionless units ($\omega\tdis$) or in physical units, is almost
independent of the initial half-mass radius, $\rhi$, when $\rhi$
relative to the initial Jacobi (or tidal) radius, $\rji$, is larger
than $\rhji(\equiv\rhi/\rji)\gtrsim 0.05$. For clusters that start
with $\rhji<0.05$, $\tdis$ scales mildly with $\rhji$, between
$\rhji^{0.55}$ and $\rhji^{0.35}$ for the range of $N$ we consider and
even flatter for larger $N$. Only clusters that start with
$\rhji\lesssim10^{-4}$ can lose half their stars before they reach the
influence of the tidal field. We find that in the tidal regime $\tdis$
is mainly determined by $N$ and the angular frequency: $\tdis\propto
N^{0.65}/\omega$, that is, what was also found by BM03 for Roche-lobe
filling, multi-mass clusters dissolving in tidal fields.

\citet{1997ApJ...474..223G} construct the ``vital" diagram of globular
clusters, which is a triangle in $M$ {\it vs.} $\rh$ space, outside
which clusters should have been destroyed. There are numerous globular
clusters with small $\rh$ outside this triangle, which the authors
denote as lucky survivors. We show that small clusters can in fact
survive.

Models that try to explain the evolution of the globular cluster mass
function (GCMF) from an initial power-law to a (universal) peaked
distribution by stellar dynamical processes are in difficulties, since
such models will always produce less dissolution in the outer parts of
galaxies. \citet{2007arXiv0704.0080M} have recently proposed a solution
to this problem by assuming that $\dmdt\propto M/\trh$ and so
$\dmdt\propto\sqrt{\rhoh}$ (equations~\ref{eq:dndt}\&\ref{eq:trh},
with $\xie$ constant), that is, their $\tdis$ is determined by
internal relaxation effects only and is independent of the strength of
the tidal field. However, our results show that $\xie$ is not constant
and, therefore, $\dmdt$ does not scale as $\sqrt{\rhoh}$.  This makes
the universality of the GCMF a problem (again), when trying to explain
this by dynamical evolution alone.

\section*{Acknowledgement}
We thank an anonymous referee for constructive comments.  We are
grateful to Douglas Heggie and Ivan King for discussions.  MG enjoyed
discussions with Henny Lamers during his stay in Santiago.  The
simulations were done on the GRAPE-6~BLX64 boards of the European
Southern Observatory in Garching.  This research was supported by the
DFG cluster of excellence Origin and Structure of the Universe
(www.universe-cluster.de).

\bibliographystyle{mn2e}

\end{document}